# Single-Frequency High-Power Continuous-Wave Oscillation at 1003 nm of an Optically Pumped Semiconductor Laser

M. Jacquemet[a], M. Domenech[a], J. Dion[b], M. Strassner[b],
G. Lucas-Leclin[a], P. Georges[a], I. Sagnes[b] and A. Garnache[c]

[a] Laboratoire Charles Fabry de l'Institut d'Optique
UMR 8501, Centre Universitaire, Bat. 503, 91403 Orsay cedex (France)
*gaelle.lucas-leclin@iota.u-psud.fr*

[b] Laboratoire de Photonique et de Nanostructures
CNRS UPR20, Route de Nozay, 91460 Marcoussis (France).

[c] Centre d'Electronique et de Micro optoélectronique de Montpellier, CNRS UMR 5507,
Universite Montpellier II, Place E. Bataillon, 34095 Montpellier cedex 5 (France).

## ABSTRACT

This work reports single-frequency laser oscillation at λ = 1003.4 nm of an optically pumped external cavity semiconductor laser. By using a gain structure bonded onto a high conductivity substrate, we demonstrate both theoretically and experimentally the strong reduction of the thermal resistance of the active semiconductor medium, resulting in a high power laser emission. The spectro-temporal dynamics of the laser is also explained. Furthermore, an intracavity frequency-doubling crystal was used to obtain a stable single-mode generation of blue (λ = 501.5 nm) with an output power around 60 mW.

## 1. INTRODUCTION

Optically pumped Vertical-External-Cavity Surface Emitting Lasers (VECSEL's) combine the approaches of diode-pumped solid-state lasers and engineered semiconductor lasers, generating both circular diffraction limited output beams and high average powers [1,2]. However the poor thermal conductivity of III-V materials might prevent an efficient heat removal from the active region, which results in an excessive heating of the semiconductor structure incompatible with high power emission. Mainly two solutions have been described in the literature: first growing the structure upside down and removing the GaAs substrate to dissipate the heat directly through the Bragg mirror [1,3]; or secondly removing the heat through the top of the semiconductor structure by bonding it to a material of high thermal conductivity and good optical quality [2,4,5]. In this work we propose solid-liquid inter-diffusion (SLID) bonding to transfer temperature sensitive components without introducing additional strains. This technique benefits from a relatively low bonding temperature (~200° C) and a good temperature stability of the bond since the melting point of the $AuIn_2$ alloy is 480° C. Further high temperature processing steps, such as plasma enhanced vapor phase deposition of dielectric materials, are thus possible without dissolving the $AuIn_2$ bond.

Output powers up to 500 mW in a single transverse and longitudinal mode have been obtained at 1003 nm from an optically pumped ½ VCSEL bonded onto a SiC heatspreader. Furthermore, intracavity second harmonic generation is reported at the blue-green wavelength of 501.5 nm, which corresponds to a hyperfine transition of the molecular iodine ($^{127}I_2$) of great interest in metrological applications as well as in high-resolution spectroscopy [6,7].

## 2. ½ - VCSEL STRUCTURE FOR THERMAL MANAGEMENT

**2.1 Gain structure design and fabrication**

The $\lambda_L$=1010 nm ½ -VCSEL designed for $\lambda_p$=808 nm pumping used in the present work was grown by MOCVD on a 350 µm – thick GaAs (100) substrate [8]. The multilayer Bragg mirror is made with 27.5 pairs of GaAs/AlAs. The 3.75 λ-thick active region consists of $N_{QW}$ = 5 compressively strained $L_{QW}$=7 nm InGaAs/GaAs quantum wells (QWs) distributed among the 7 optical standing-wave antinode positions with a distribution function 1-1-1-0-1-0-1 (starting from the top surface) such as the carrier density remains almost equal in all QWs. We have optimized the number of QWs to obtain a low laser threshold and a large differential gain at 300 K with 1% of total optical cavity losses [9]. The low number of QWs, as compared to other published designs, avoids the need to grow strain-compensating layers between the QWs [1-4]. On top of the gain region a 30 nm AlAs confinement layer and an 8 nm caping layer were added.

A second structure, with exactly the same layers but an AlGaAs etch-stop layer between the GaAs substrate and the Bragg mirror, was grown on reverse order on a GaAs substrate for further processing. This ½-VCSEL was bonded onto a 280-µm thick Si substrate (thermal conductivity of $\kappa_{s1}$ =150 W/m/K) on the Bragg mirror side by solid-liquid inter-diffusion (SLID) bonding with $AuIn_2$ (figure 1) [10]. A 30 nm-thick Ti layer was first evaporated on both the ½ -VCSEL and the SiC heatspreader to improve adhesion and to prevent diffusion. Next, a 150 nm –thick Au layer was deposited on the SiC substrate. On top of the active structure we added Au (150 nm) and In (600 nm) layers; then a 20 nm-thick Au layer was deposited to prevent the oxidation of the Indium layer prior to bonding. The bonding temperature was 200° C, which was a trade-off between the diffusion speed and the strain introduced by the bonding process due to the different thermal expansion coefficients of both materials. A pressure of about 250 kg / cm² was applied to provide a good surface contact of the two wafers while the temperature was ramped to 200° C and maintained at this temperature for two hours. The pressure was kept constant during the cooling-down process. The final thickness of the $AuIn_2$ bonding layer was found to be 0.9 µm. Then both mechanical polishing and wet selective etching at room temperature was utilized to remove the GaAs substrate. Finally, a $Si_3N_4$ anti-reflection coating was deposited by plasma enhanced chemical vapor phase deposition (PECVD) at high temperature (280° C) to protect the top surface from oxidation and to weaken the residual etalon resonances of the semiconductor structure.

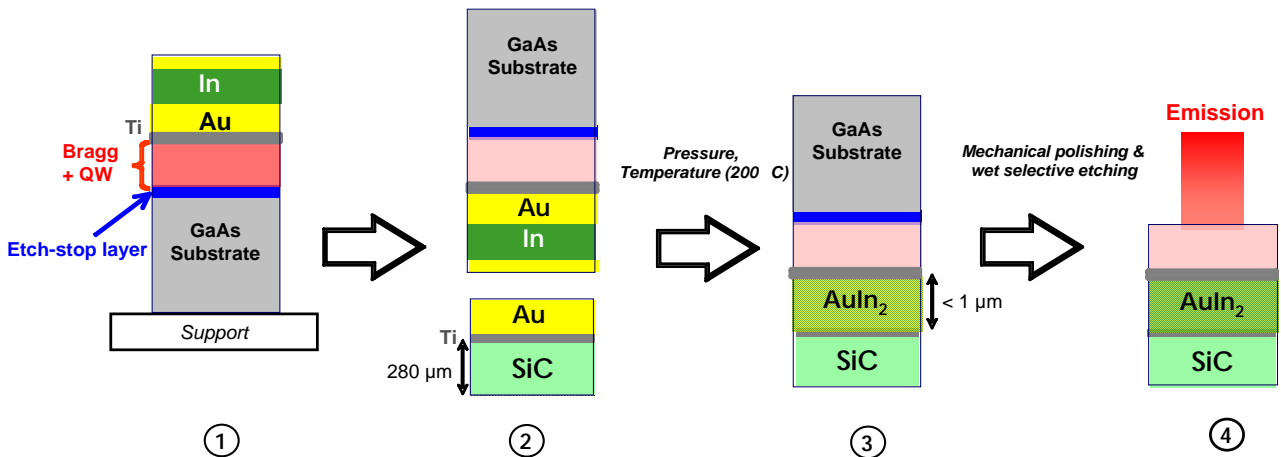

Figure 1: Description of the SLID bonding process of the ½ VCSEL structure onto a SiC heatspreader.

The reflectivity and photoluminescence (PL) spectra of the bonded structure are shown in Figure 2, demonstrating an optimized design with a good matching between the central wavelength of the Bragg mirror and the QW gain. Neither a PL broadening nor a PL wavelength shift have been observed for the bonded structure, demonstrating that our $AuIn_2$ bonding technique does not modify the optical properties of the QWs. With a residual air-structure interface reflection of about 6%, and a calculated pump absorption at $\lambda_P = 808$ nm of nearly 70 % in the barrier layers, the fraction of incident pump power absorbed in the active region is evaluated to $A_p=66\%$. The remaining 30% pump power is absorbed in the GaAs layers of the Bragg mirror.

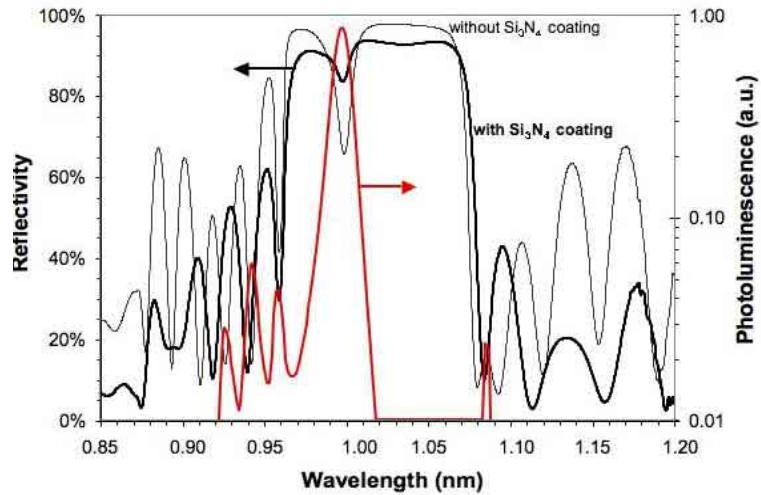

Figure 2: Reflectivity of the half-VCSEL bonded onto the SiC substrate (left; black curve : with AR coating, grey curve : without AR coating) and photoluminescence spectrum (red curve, right) under low-power excitation before the AR coating deposition. The residual etalon effect of the ½ -VCSEL remains visible on the reflectivity spectrum around 1.0 µm.

**2.2 Theoretical and experimental evaluation of the thermal properties of the gain structures**

The heat diffusion in the whole semiconductor structure has been theoretically studied assuming a perfect metal bonding between the Bragg mirror and the SiC substrate. The pump absorption was supposed to decrease exponentially through the active region and the Bragg mirror; the pump beam transverse profile was top-hat. We have evaluated the average thermal resistance $R_{th}$ of the ½ -VCSELs over the laser beam area, measured between the top surface of the Peltier cooler and the QWs. The values of $R_{th}$ are given relatively to the incident pump power. Assuming an internal quantum efficiency $\eta_i$ of 80% for the QWs and of 10% for the GaAs in the Bragg mirror, 50% of the incident pump power is converted to heat.

    On the one hand, we used an analytical three-layer model (active region/susbtrate/copper block) in which the substrate itself is perfectly bonded onto an infinite copper block (thermal conductivity of 4 W/m/K) [11]. The active region consists of the QWs and the Bragg mirror layers, with a total thickness $e_v=5.45$µm. A mean value of $\kappa_v=32$ W/m/K was considered as its thermal conductivity. We have evaluated that for pump beam waist radii larger than the critical value $r_c= e_v\,(\kappa_s/\kappa_v)$, the major contribution to the temperature increase in the QW layers arises from the substrate. Then the heat flow in the ½–VCSEL is purely 1D, and the temperature rise only depends on the pump power *density*. These critical pump radii are $r_c= 7.7$ µm and 83.4 µm for the GaAs and SiC substrates respectively. On the other hand, the thermal resistance has been evaluated with a finite-element numerical simulation (FEMLAB). A 10-mm lateral width of the sample was considered. Furthermore, a 50 µm-thick heat-paste layer (thermal conductivity of 2W/K/m) embedded between the substrate and copper block and the substrate could be taken into account The thermal resistances evaluated from both the numerical and the analytical models are compared in Figure 3. The analytical model, which mainly neglects the heat-paste layer contribution, slightly underestimates the thermal resistance from 2 to 60% in the range of pump radius investigated here (30-450µm). It is thus a simple and efficient tool for the evaluation of the thermal resistance of the structures.

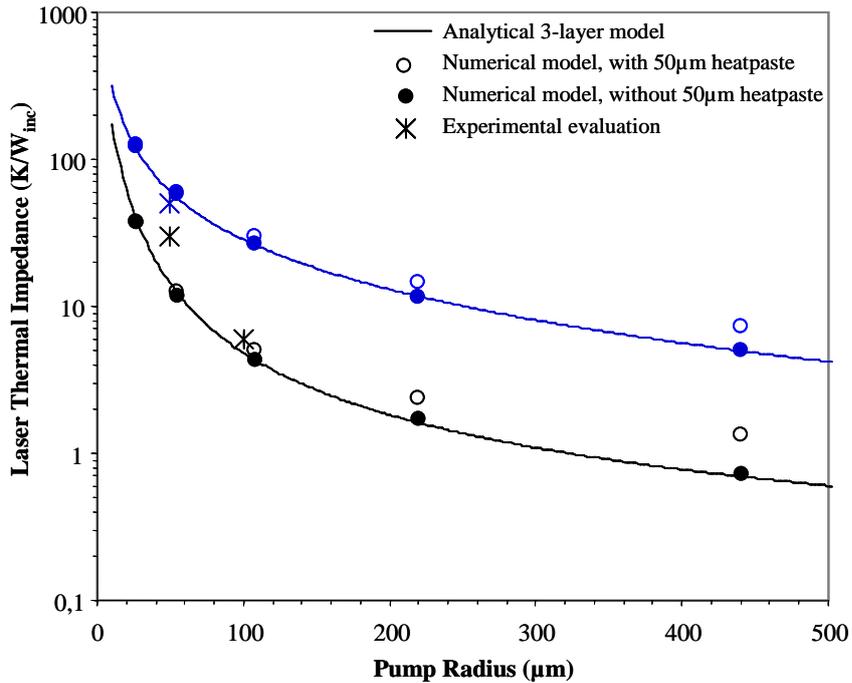

Figure 3: Theoretical evaluations (lines: analytical, circles: numerical) of the thermal resistance of the ½-VCSEL structures on respectively SiC and GaAs substrates, vs the pump radius, and comparison with the experimental values (crosses).

These simulations have been compared to the experimental evaluation of the thermal resistance of the two samples under laser operation close to threshold, through the measurement of the emission wavelength shift with the incident pump power and the heatsink temperature. For a pump waist radius of 100 µm, a thermal resistance of 6 $K/W_{inc}$ was measured for the structure bonded onto SiC, which is in relatively good accordance with the theoretical value of 5.6 $K/W_{inc}$. This value is much lower than the thermal resistance measured under the same experimental conditions for the structure on the GaAs substrate, which was about 50 $K/W_{inc}$. Finally, the temperature at the top surface of the ½-VCSELs has been measured under cw optical pumping in the spontaneous emission regime (no laser cavity) with an infrared camera operating in the 8-12µm range [12]. The lateral resolution of the mapping was 60µm, thus only suited for large pump beam waists. With a maximum temperature rise limited to 15°C for a 4 W pump beam focused on a 250 µm radius waist, the structure bonded onto SiC obviously demonstrates the lower thermal resistance (figure 4). Furthermore these maps show the good homogeneity of our bonding process.

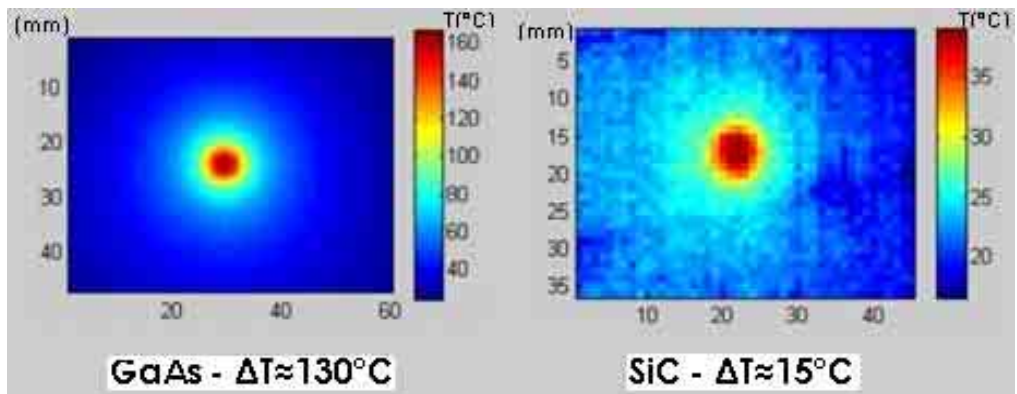

Figure 4: Temperature maps of the ½-VCSEL structures under optical pumping ($\lambda$ = 808 nm, $P_p$ = 4 W, $W_p$ = 250 µm) without laser operation.

# III. INFRARED LASER CHARACTERIZATION

## 3.1 Experimental set-up

The VECSEL system was set up in a simple and compact cavity configuration as it is shown in Figure 5. The half-VCSEL was mounted onto a copper heatsink to control the temperature of the device with a Peltier module. The laser was pumped with a continuous wave (cw) fiber-coupled $\lambda_p = 808$ nm laser diode delivering up to 7 W focused on approximately a 100 µm radius spot with an incidence angle of 20°. The external cavity was simply formed by the VCSEL and a concave dielectric mirror with a reflectivity of $R_{oc}= 98.8$ % at $\lambda_L=1010$ nm and a curvature radius of 100 mm. The overall cavity length was around $L_c=95$ mm, leading to a cavity Free-Spectral-Range (FSR) of ~ 5 pm.

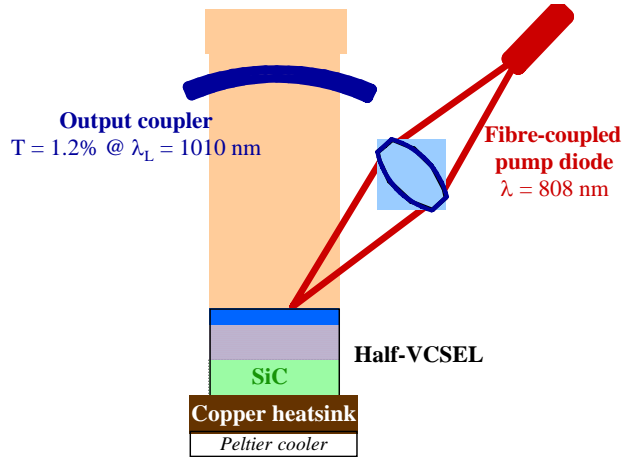

Figure 5: Experimental set-up for high-power laser emission in the infrared.

## 3.2 High power laser characterization: experiment and theory

In the compact two-mirror cavity and without any intracavity element, an output power of 1.7 W at the maximum incident pump power of 7 W was obtained with the ½ - VCSEL bonded on SiC at 283 K. On the contrary, no laser operation was achievable with the structure on GaAs in the same cavity and pump conditions, and the maximum output power was only 100 mW for a pump waist radius of 50 µm. This demonstrates once more the effectiveness of the bonding technique on a high conductivity substrate for high power laser emission. Further laser investigations will thus only be reported with the ½ - VCSEL bonded onto SiC: the incident pump power at threshold was 0.8 W in cw operation at 283 K, corresponding to an incident pump density of $I_{inc}^{th}=2.2$ kW/cm$^2$, and the CW slope efficiency was about 32 %. The laser oscillation occurs around 1009 nm at the maximum pump power, with a cw spectral bandwidth of nearly 2 nm. We believe that this linewidth is limited by the QW's gain jitter induced by the pump power fluctuations.

These experimental results have been compared to a theoretical evaluation of the VECSEL behavior based on the following equations:

$$I_{inc} = J_{current} \times \frac{N_{QW} hc}{\lambda_p A_p e} \qquad (Eq.1)$$

$$g_{th} = g_0 \ln \frac{J_{current}^{th}}{J_{tr}} \approx \frac{T_i + (1 - R_{oc})}{2\Gamma_c N_{QW} L_{QW}} \qquad (Eq.2)$$

$$\eta_e \approx \eta_i \frac{\lambda_p}{\lambda_L} A_p \times \frac{1 - R_{oc}}{(1 - R_{oc}) + T_i} \qquad (Eq.3)$$

Equations (Eq. 1) and (Eq. 2) respectively sets the relation linking the incident pump power density $I_{inc}$ to the current density per QW $J_{current}$ (in A/cm$^2$), and the material gain at threshold $g_{th}$ to the cavity parameters. The values for $g_0$ ($1200$ cm$^{-1}$) and the $J_{tr}$ ($50$ A/cm$^2$), which make the connection between the maximal material gain and the current density per QW, are deduced from reference 9 for InGaAs/GaAs quantum wells. The longitudinal confinement factor $\Gamma_c$ is 2 and the intracavity losses $T_i$ are evaluated to 0.2%. The external quantum efficiency $\eta_e$ at the laser output, assuming a pump-to-laser beam recovering integral close to unity, is given by (Eq. 3). Finally, the evolution of the threshold with temperature is theoretically taken into account through the characteristic temperature $T_0$, evaluated to 60 K in these experiments. Figure 6 shows the very good agreement between the theoretical evaluation of the output power with the incident pump power, and our experimental results. An actual output power limited to 1.8 W is predicted for an incident pump power of 7.9 W and a 100 µm radius pump waist.

These simulations have been used to estimate the maximum output power which could be extracted from our ½-VCSEL at T = 283 K, with an optimization of both the incident pump power and the pump beam waist. In these calculations the thermal resistance of the structure was evaluated with a 50 µm-thick heatpaste layer between the copper heatsink and the substrate (see figure 3). Obviously the low thermal conductivity of GaAs limits the available laser power to 0.1 W (figure 6, inset). On the contrary an output power as high as 6.7 W should be obtained with a 30 W incident pump beam focused onto a 260 µm radius waist with the structure bonded on SiC.

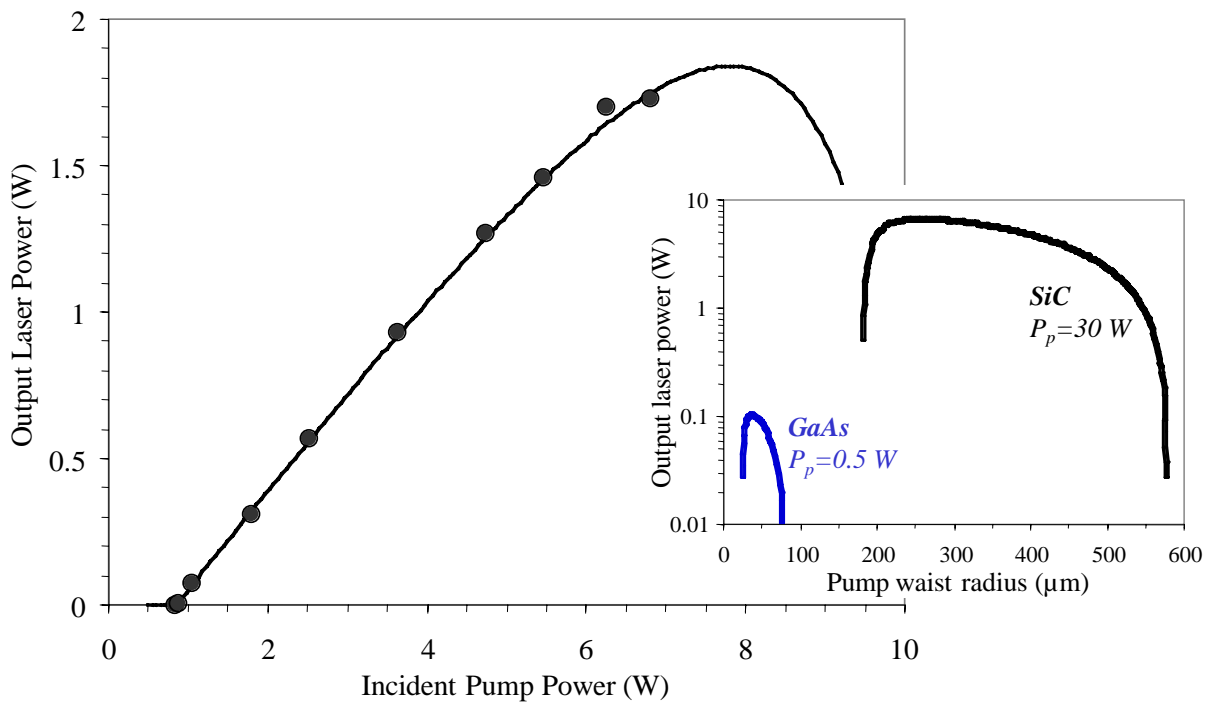

Figure 6: Experimental (full circles) and theoretical (line) output laser emission from the structures bonded onto SiC at T = 10°C, for a pump beam waist radius $W_p$ = 100 µm. Inset : theoretical optimization of the output laser power vs the pump waist radius for the ½ -VCSEL on GaAs and the one bonded onto SiC ; the incident pump power is set for maximum laser emission of each structure; T = 10°C, output coupler T = 1.2%.

### 3.3. Single-frequency laser operation

The insertion of a solid 50 µm-thick low-finesse etalon in the compact cavity set-up described in figure 5 results in a narrow-line laser emission tunable from ~ 998 nm to ~ 1010 nm by rotating the etalon and changing the pump position on the ½ -VCSEL structure. The maximum output power was 1.3 W at 1006.5 nm for an incident pump power of 7 W; the reduction of the output power by about 30 % as compared to the previous results is due to the additional losses introduced by the solid etalon. On the entire incident pump power range the emission wavelength remains strongly controlled by the etalon with a spectral shift less than 0.2 nm. The beam quality factor $M^2$ relating the spatial beam properties to an ideal Gaussian beam was about 1.3 at the highest pump power. The linear polarization of the laser beam was aligned along the [110] semiconductor crystal axis, due to a slight gain dichroism between [110] and [1-10] crystal axis.

A true single-frequency operation was obtained at 1003 nm for an output power limited to 0.5 W at the maximum incident pump power. That further decrease of the output power can be mainly attributed to the alignments of the cavity and the etalon to force a stable single-frequency laser line. The single-frequency laser spectrum was analyzed with a 50 mm confocal scanning Fabry-Perot interferometer with a FSR of around 1.5 GHz (see inset in figure 7). The side-mode suppression ratio is at least 20 dB, but the scanning Fabry-Perot apparatus function limits the measurement of the laser linewidth. However in this configuration, the laser linewidth should be limited by acoustic and/or thermal noise at the kHz level [4,13]. No stabilization of the laser emission frequency has been carried out.

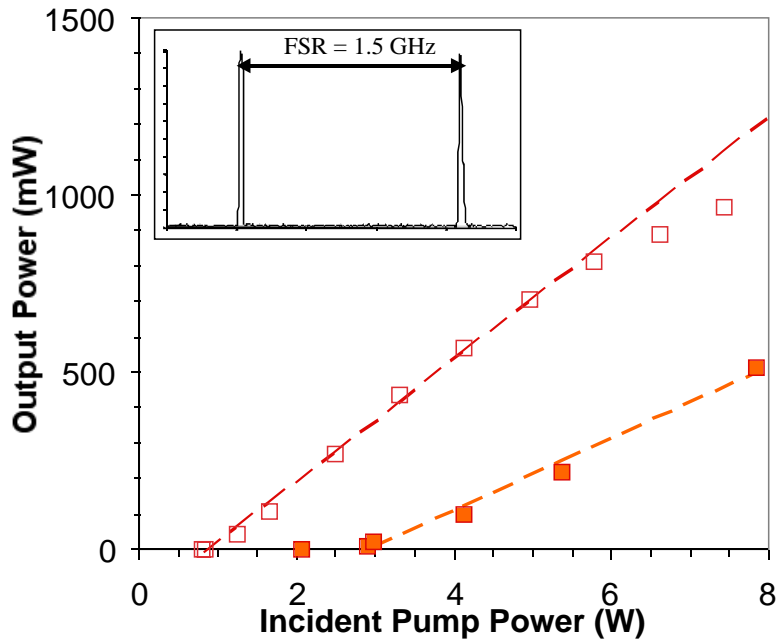

Figure 7: CW laser output power vs the incident pump power with the intracavity 50 µm-thick etalon for the structure bonded onto SiC; open squares: narrow spectrum laser emission at 1003 nm; full squares: true single-frequency operation at 1003 nm; inset: scanning Fabry-Perot trace for the emission at 1003.4 nm demonstrating single-frequency operation with the FP etalon inside the laser cavity. The ½-VCSEL temperature is 10°C.

It has recently been demonstrated both theoretically and experimentally that a single-frequency operation is achievable in short cavity ($L_c < 100$ mm) VECSEL without any intracavity filter, thanks to their ideal multimode homogeneous gain and the fact that non-linear mode coupling is negligible [13,14]. Indeed, though at the very beginning of the laser operation several thousands of longitudinal cavity modes are amplified, the laser collapses to a single mode after a characteristic time $t_c$. But the emission would remain multimode if any process (gain or mirror jitter, thermal fluctuations…) disturbs the laser dynamics on a time scale shorter than $t_c$. The characteristic time $t_c$ necessary to achieve a single-mode laser operation after a *strong* perturbation arises (100% gain modulation e.g.) is equal to the time $t_g$ within which the non-stationary laser bandwidth $\Delta_L$, given by (Eq. 4), is equal to one FSR of the laser cavity:

$$\Delta_L = \Gamma_{g,f} \sqrt{\frac{\ln 2}{\gamma_0 t_g}} \qquad \text{(Eq.4)}$$

In this equation, $t_g$ is the generation time elapsed since the beginning of the laser emission, $\gamma_0$ is the broadband cavity loss rate (proportional to $1/L_c$; $\gamma_0 = 22.1 \times 10^6 \text{ s}^{-1}$ in our operating conditions), and $\Gamma_{g,f}$ is the gain or filter bandwidth, depending on the intracavity elements [15]. The single-mode characteristic time $t_c$ is plotted in figure 8 as a function of the cavity length, with ($\Gamma_f = 1.1$ nm FWHM – parabolic fit) and without ($\Gamma_g = 34$ nm) the 50-µm Fabry-Perot etalon inserted.

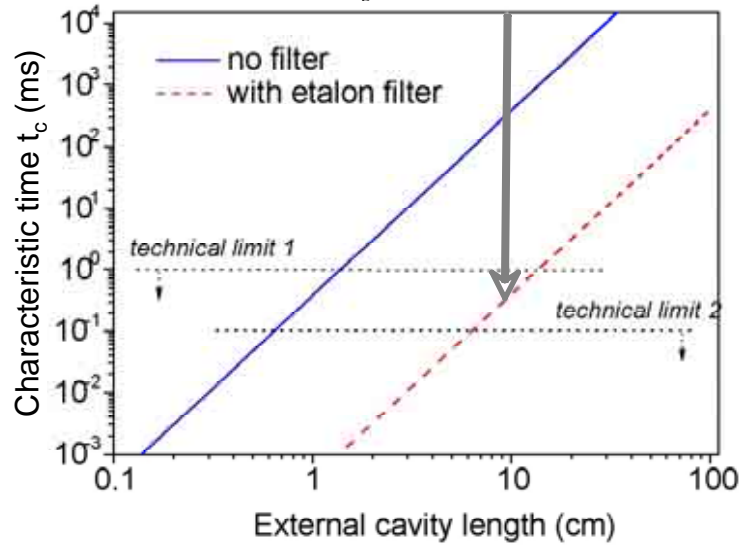

Figure 8: Characteristic generation time $t_c$ for an ideal homogeneous laser to collapse to single-mode operation, with and without a 50µm thick etalon filter in the cavity; non-linear mode coupling is neglected. Dot line #1: acoustic or/and thermal noise typical perturbation time; dot line #2: multimode pump fluctuations perturbation time. The top arrow shows the actual cavity length used in the experiment.

In our experimental conditions, without the etalon filter inside the laser cavity, the laser emission was not single-mode and the cw (stationary) laser bandwidth $\Delta_L$ covered about several hundreds cavity modes. From (Eq. 4) it corresponds to a generation time - the duration necessary for the laser spectrum to narrow into a single mode before a perturbation arises - of about 100 µs. Since this value is one order of magnitude shorter than the typical characteristic time of acoustic or/and thermal noise (in the millisecond range [3, 13]), we believe that the multimode nature of our fiber-coupled pump laser diode generates strong pump intensity fluctuations at a time scale shorter than 100 µs, which in turns destabilize the central wavelength of the gain spectrum and interrupt the spectral narrowing of the VECSEL. From the simulations in figure 8 it is obvious that even our low-finesse etalon, with a FSR of 2 THz ($\Delta\lambda \sim 7$ nm @ 1003 nm) which covers numerous longitudinal modes, helps to reduce the characteristic time of the single-mode operation below the typical fluctuations time of the set-up induced by acoustic, thermal noise and/or pump-induced gain fluctuations. It allows thus to obtain a strong single-mode operation, as observed experimentally.

## VI. INTRACAVITY SECOND HARMONIC GENERATION AT 501.7 NM

A four-mirror Z-cavity with a total length of around 630 mm was implemented for intracavity second harmonic generation at 501.7 nm (figure 9). Mirrors $M_1$ and $M_2$ had a radius of curvature of 200 mm, and a reflectivity > 99.98 % between 900-1100 nm. Mirror $M_3$ was a 75 mm spherical mirror with a reflectivity > 99.9 % in the range 650-1100 nm. The laser cavity has two different outputs with similar powers at 501.7 nm because of the high transmission of all the mirrors in the visible range. A type-I temperature-tuned non-critical phase-matching $KNbO_3$ crystal was placed inside an oven at a temperature of 75.6 °C. The nonlinear crystal was *b*-cut, 9.5 mm-long and anti-reflection coated for both fundamental and second harmonic wavelengths. The waist radius inside the $KNbO_3$ crystal was evaluated to about 90 µm. A Lyot filter and a 100 µm-thick FP etalon were also inserted not only to select the desire emission wavelength, but also to obtain a stable single frequency operation at both wavelengths 1003.4 nm and 501.7 nm, which was impossible to achieve without the etalon. The crystal was oriented such as the intracavity fundamental laser beam at 1003.4 nm was polarized along the *a*-axis of the non-linear crystal, while the second harmonic generated beam was polarized along the *c*-axis.

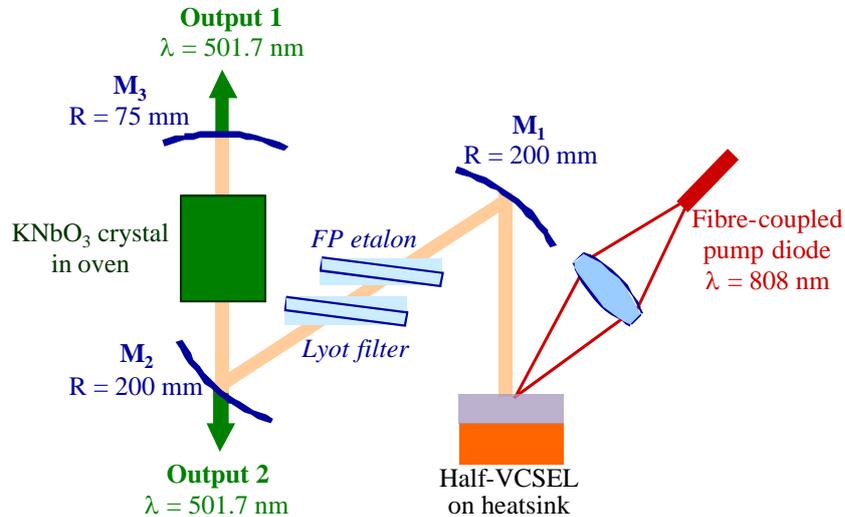

Figure 9: Experimental set-up achieved for the second harmonic generation process at 501.7 nm in a four- mirror cavity with the Lyot Filter, the etalon and the oven for the KNbO$_3$ doubling crystal.

Figure 10 represents the single-frequency output power of the green-blue emission as function of the incident pump power at 808 nm. A total maximum cw output power of 62 mW at 501.7 nm in two beams for a 6.5 W pump power was obtained. The infrared intracavity power was estimated to be about 10 W, leading to a second harmonic generation conversion efficiency of only 0.6 %. This low conversion efficiency can be attributed to the constraints imposed by the oven dimensions, which lead to a rather large waist inside the crystal and a confocal parameter much longer (~ 44 mm) than the crystal length [15]. Moreover, the residual losses introduced by the intracavity elements and the mirrors reduce the available infrared laser power. The frequency spectrum of the fundamental infrared laser beam was also analysed with the same Fabry-Perot interferometer as previously. The inset of figure 10 shows a typical scanning trace, which demonstrates the single-frequency operation of the laser. Though these results are very preliminary and the second harmonic generation process is not yet fully optimized as explained, the reached single frequency output power at 501.7 nm is similar to the one obtained with an Yb-doped laser crystal and a KNbO$_3$ frequency-doubling crystal by Jacquemet et al. [16], with the advantage of a simpler design of the laser cavity. Therefore, considering these as previous measurements, future prospects focused on the enhancement of the SHG, on the stabilization of the laser emission and on the change of the doubling crystal are going to be carried out.

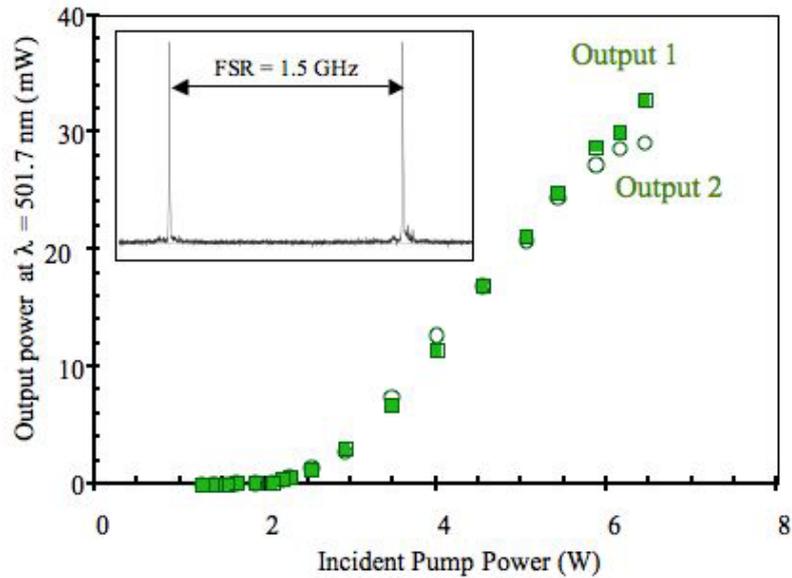

Figure 10: CW laser output power at λ= 501.7 nm vs the incident pump for a 75.6 °C $KNbO_3$ crystal temperature. The inset shows the frequency spectrum confirming single-frequency behavior in the blue-green.

## VII. CONCLUSION

In conclusion, in this work the highest output power in a single-frequency operation of a diode-pumped VECSEL has been reported. Thanks to an original technique based on the $AuIn_2$ solid liquid inter diffusion bonding, the active structure has been bonded on a SiC substrate exhibiting a low thermal impedance. The $AuIn_2$ bonding allows further processing steps at elevated temperatures and does not induce any strain-related degradation of the optical properties of the structure. We then obtained a maximum output power of 1.7 W in a simple and compact plano-concave cavity, limited by the available incident pump power. The improvement of the thermal dissipation from the active region has been experimentally checked with both laser characterizations and temperature mappings under high power optical pumping. It was theoretically explained through the evaluation of the thermal resistance of the ½-VCSEL, which allows us to predict the conditions for optimal laser operation.

With the insertion of only a solid Fabry-Perot etalon it has been possible to pick the emission wavelength at 1003 nm and to force a single-frequency operation with up to 500 mW of output power. A simple model has been used to explain that laser emission on a single longitudinal mode is possible even with our low finesse and large FSR solid etalon. We further demonstrate that, due to the multimode nature of the optical pump source used, an intracavity filter is necessary to develop high power single-frequency VECSEL systems.

Finally, motivated by the interest of blue-green single-frequency emission around 501.7 nm for metrological applications such as iodine-based laser stabilization, second harmonic generation had been performed with a $KNbO_3$ nonlinear crystal. A maximum visible power of 62 mW has been generated in a four-mirror Z-cavity. These results can be compared to those obtained with bulk ytterbium-doped crystals in a more complicated architecture.

## REFERENCES

1. M. Kuznetsov, F. Hakimi, R. Sprague and A. Mooradian, "Design and characteristics of high-power (>0.5-W CW) diode-pumped vertical-external-cavity surface-emitting semiconductor lasers with circular $TEM_{00}$ beams," IEEE J. Selec. Topics in Quantum Electron. 5, 561 (1999).

2. W. J. Alford, T. D. Raymond and A. A. Allerman, "High power and good beam quality at 980 nm from a vertical external-cavity surface-emitting laser," J. Opt. Soc. Am. B. 19, 663 (2002).
3. S. Lutgen, T. Albrecht, P. Brick, W. Reill, J. Luft and W. Späth, "8-W high-efficicency continuous-wave semiconductor disk laser at 1000 nm," App Phys Lett 82, pp 3620-3622 (2003)
4. R. H. Abram, K. S. Gardner, E. Riis and A. I. Ferguson, "Narrow linewidth operation of a tunable optically pumped semiconductor laser," Opt. Exp. 12, 5434 (2004).
5. J. E. Hastie, J. M. Hopkins, S. Calvez, C. W. Jeon, D. Burns, R. H. Abram, E. Riis, A. I. Ferguson and M. D. Dawson, "0.5-W single transverse-mode operation of an 850 nm diode-pumped surface-emitting semiconductor laser," IEEE Photon. Technol. Lett. 13, 894 (2003).
6. F. Du Burck, C. Daussy, A. Amy-Klein, A. Goncharov, O. Lopez, and C. Chardonnet, "Frequency Measurement of an Ar+ Laser Stabilized on Narrow Lines of Molecular Iodine at 501.7 nm," IEEE J.Trans. Intrum. and Meas., 54, 754-758 (2005)
7. W. Y. Cheng, L. Chen, T. H. Yoon, J. L. Hall and Y. Ye, "Sub-doppler molecular-iodine transitions near dissociation limit (523-498 nm)," Opt. Lett. 27, 571 (2002).
8. J. Dion, I. Sagnes and M. Strassner, "High output power GaAs-based Vertical External Cavity Surface Emitting Lasers achieved by AuIn2 Solid Liquid Inter Diffusion Bonding", State-of-the-Art Program on Compound Semiconductors XLI (2004)
9. L.A.Coldren, S.W. Morzine, Diode lasers and photonic integrated circuits, chapter IV, 1995, Wiley Interscience
10. L. Bernstein, ``Semiconductor joining by SLID Process, I. The systems Ag-In, Au-In and Cu-In," J. Electrochem. Soc., **113**, 1282-1288 (1966)
11. M.Reichling and H. Grönbeck, "Harmonic heat flow in isotropic layered systems and its use for thin film thermal conductivity measurements", J.Appl.Phys. 75, 1914-1922 (1994).
12. S. Chénais, S. Forget, F. Druon, F. Balembois and P. Georges, "Direct and absolute temperature mapping and heat transfer measurements in diode-end-pumped Yb:YAG", Appl.Phys.,B 79 p.221 ( 2004)
13. A. Ouvrard, A. Garnache, L. Cerutti, F. Genty, D. Romanini, *"Single Frequency Tunable Sb–based VCSELs emitting at 2.3 μm"*, IEEE Photon. Techn. Lett.,17, 2020-2022 (2005)
14. A. Garnache, A. Ouvrard and D. Romanini, "Spectro-temporal dynamics of external-cavity VCSELs: Single-Frequency operation" Proc. IEEE CLEO Europe 2005, paper CB19
15. A. Garnache, A. Kachanov, F. Stoeckel and R. Houdre, "Diode-pumped broadband vertical external cavity surface emitting semiconductor laser applied to high-sensitivity intracavity absorption spectroscopy," J. Opt. Soc. Am. B 7, 1589 (1999).
16. M. Jacquemet, F. Druon, F. Balembois, P. Georges, and B. Ferrand, "Blue-green single-frequency laser based on intracavity frequency doubling of a diode-pumped Ytterbium-doped laser," Optics Express 13, 2345-2350, (2005).